# Multi-scale strain-stiffening of semiflexible bundle networks

Izabela K. Piechocka,[†ab] Karin A. Jansen,[†a] Chase P. Broedersz,[cd] Nicholas A. Kurniawan,[*ae] Fred C. MacKintosh[*f] and Gijsje H. Koenderink[*a]

[a]*FOM Institute AMOLF, 1098 XG Amsterdam, The Netherlands.*
[b]*ICFO-Institut de Ciencies Fotoniques, 08860 Castelldefels (Barcelona), Spain.*
[c]*Lewis-Sigler Institute for Integrative Genomics and the Department of Physics, Princeton University, Princeton, NJ 08540, USA.*
[d]*Arnold-Sommerfeld-Center for Theoretical Physics and Center for NanoScience, Ludwig-Maximilians-Universität München, Theresienstrasse 37, D-80333 München, Germany.*
[e]*Department of Biomedical Engineering, Eindhoven University of Technology, P.O. Box 513, 5600 MB Eindhoven, The Netherlands.*
[f]*Department of Physics and Astronomy, VU University, 1081 HV Amsterdam, The Netherlands.*
[†]I. K. Piechocka and K. A. Jansen made equal contributions to this paper.
*To whom correspondence should be addressed: N. A. Kurniawan (E-mail: kurniawan@tue.nl); F. C. MacKintosh (E-mail: fcm@nat.vu.nl); G. H. Koenderink (E-mail: gkoenderink@amolf.nl).

## ABSTRACT

Bundles of polymer filaments are responsible for the rich and unique mechanical behaviors of many biomaterials, including cells and extracellular matrices. In fibrin biopolymers, whose nonlinear elastic properties are crucial for normal blood clotting, protofibrils self-assemble and bundle to form networks of semiflexible fibers. Here we show that the extraordinary strain-stiffening response of fibrin networks is a direct reflection of the hierarchical architecture of the fibrin fibers. We measure the rheology of networks of unbundled protofibrils and find excellent agreement with an affine model of extensible wormlike polymers. By direct comparison with these data, we show that physiological fibrin networks composed of thick fibers can be modeled as networks of tight protofibril bundles. We demonstrate that the tightness of coupling between protofibrils in the fibers can be tuned by the degree of enzymatic intermolecular crosslinking by the coagulation Factor XIII. Furthermore, at high stress, the protofibrils contribute independently to the network elasticity, which may reflect a decoupling of the tight bundle structure. The hierarchical architecture of fibrin fibers can thus account for the nonlinearity and enormous elastic resilience characteristic of blood clots.

# INTRODUCTION

Polymer bundles are found everywhere in Nature. Inside cells, polymer bundles are present as part of the cytoskeleton, which is a space-spanning composite network made up of actin filaments, microtubules and intermediate filaments.[1,2] Actin and intermediate filaments can be classified as semiflexible polymers, meaning that their thermal persistence length is comparable to their contour length,[2,3] whereas microtubules are often considered as rigid rods.[4] A large number of accessory proteins such as molecular motors and crosslink proteins organize these polymers into higher-order structures tailored for specific tasks, including bundles that act as reinforcing or force-generating elements.[5-7] Polymer bundles also form the main structural element of the extracellular matrix in connective tissues. However, contrary to cytoskeletal proteins, extracellular matrix proteins can spontaneously form bundled fibers without the need for accessory cross-linker proteins. Collagen I for instance self-assembles into rope-like, axially ordered bundles that endow tissues with a large tensile strength,[8] whereas the plasma protein fibrin forms axially ordered bundles that reinforce blood clots.[9]

Semiflexible polymer bundles have recently started to raise a lot of theoretical attention because their hierarchical structure endows them with unique mechanical properties. The molecular packing geometry of biopolymer bundles is generally governed by an energetic trade-off between filament twisting and interfibril adhesion[10-12]. The bending stiffness of these bundles is highly tunable, being sensitive to the intrinsic properties of the number of constituent polymers, their intrinsic mechanical properties and the strength of coupling among them.[13] These bundle properties have begun to be exploited in materials science, as exemplified by fibers made of carbon nanotubes[14,15] and responsive gels from designer supramolecular polymers.[16,17]

Theoretical models have been developed specifically to address the molecular basis of the structure and linear elasticity of bundles of actin filaments bridged at discrete binding sites by crosslinking proteins.[18-22] By contrast, much less is known about the molecular mechanisms governing the mechanical properties of fibrin and collagen bundles. They tend to be much larger in size compared to actin bundles, involving hundreds or even thousands of subunits[23,24] compared to tens of subunits[5] per cross-section in case of actin. They have a more complex molecular packing structure and it is less clear how the subunits are held together in the bundle than in the case of actin bundles. Moreover, fibrin and collagen are less well-ordered than actin bundles. Actin bundles generally involve a well-ordered hexagonal packing,[25,26] while both collagen and fibrin bundles are paracrystalline with long-range molecular packing order along the fiber axis but only short-range order in cross-section.[23,24,27] Moreover, the nonlinear elastic properties of these bundles and their networks remain poorly understood.[28-33]

In this work we focus on the mechanical properties of fibrin bundles. The soluble precursor of fibrin bundles is the protein fibrinogen, which circulates in plasma at a concentration of 2–3 mg/ml.[34] Fibrinogen is an S-shaped hexamer comprising two sets of three polypeptide chains, referred to as Aα, Bβ and γ.[35] Polymerization is initiated by the enzyme thrombin, which cleaves off two protective fibrinopeptides (FpA and FpB), exposing so-called A- and B-knobs. The activated fibrin monomers spontaneously assemble into polymer bundles by a two-step process. In the first step, cleavage of FpA initiates the formation of double-stranded protofibrils.[34] This is encoded in non-covalent interactions of the A- and B-knobs with complementary a- and b-holes of adjacent fibrin molecules. In the second step, cleavage of FpB promotes lateral association of the protofibrils into fibers comprising tens to hundreds of protofibrils.[36] This lateral association is promoted through B:b knob–hole interactions as well as through interactions of the long and flexible αC-regions that project out from the surface of adjacent protofibrils.[34,37] The enzyme Factor XIII (FXIII) catalyzes the formation of covalent crosslinks between the α- and γ-chains of the fibrin molecules, thus inducing a closer packing of the protofibrils in the fibers.[38]

Single-fiber stretching experiments by atomic force microscopy (AFM) have revealed that fibrin fibers have a low elastic modulus at small strain but stiffen when strained.[39,40] Moreover, fibrin fibers are elastomeric, exhibiting a remarkably large breakage strain that exceeds 200%.[39-41] Mechanical measurements on networks of fibrin fibers have revealed that fibrin also stiffens at the network level when subjected to a shear or tensile deformation.[9,29,42] This strain-stiffening response protects fibrin networks against damage from the shear stresses exerted by flowing blood and traction forces exerted by cells. Given the complex hierarchical structure of fibrin, it has been difficult to dissect the contribution of the molecular, fiber and network structure to the overall mechanical response.[43]

Recently, we proposed that by modeling fibrin fibers as bundles of semiflexible polymers, it does become possible to systematically trace the contribution of each hierarchical level of structure to the mechanical properties of fibrin.[9] However, an experimental difficulty in validating this model is that, unlike actin bundles, fibrin fibers cannot be taken apart into their constituent protofibrils and linkers, since bundling is an intrinsic property of the protofibrils. Here we show that the properties of the bundles can nevertheless be dissected by comparing the mechanical properties of fibrin networks prepared with different levels of bundling. To modulate the degree of bundling, we exploit the known sensitivity of fibrin polymerization to salt and pH conditions,[44-46] resulting in networks with bundle numbers that range over more than two orders of magnitude (2–370 protofibrils per bundle). We demonstrate that the nonlinear rheology of networks close to the protofibril (unbundled) limit is in excellent quantitative agreement with theoretical predictions for networks of semiflexible polymers, allowing us to extract the thermal persistence length and enthalpic stretch modulus of protofibrils. We next show that the mechanics of networks of fibers can be quantitatively explained by modeling the fibers as protofibril bundles. Furthermore, we find that the coupling strength between the protofibrils can be tuned by FXIII-mediated molecular crosslinking. Our findings validate the bundle model for fibrin bundles, which gives a powerful framework to integrate the mechanical properties of fibrin on different scales. Moreover, this framework is more generally applicable to other natural as well as bio-inspired fibrous materials.

## THEORETICAL BACKGROUND

We have previously shown by optical tweezers microrheology that fibrin fibers exhibit transverse thermal fluctuations with a dependence on frequency, $\omega$, that shows an $\omega^{3/4}$ power-law scaling regime above 1 kHz, which is characteristic of semiflexible polymers.[9] This observation implies that the elasticity of fibrin networks is entropic in origin and can be described by entropic models for semiflexible polymers. These models approximate a polymer by a smooth linear contour that resists bending with a quantity $\kappa$ called the bending modulus.[3] The rigidity of semiflexible polymers can be quantified by the persistence length $l_p = \kappa / k_B T$, which represents the decay length of angular correlations along the polymer contour. Here $k_B$ is Boltzmann's constant and $T$ is temperature. A polymer is called semiflexible when $l_p$ is comparable to its contour length. Because semiflexible polymers bend in response to thermal forces, their response to an applied pulling force is entropic in origin: pulling straightens out the thermally-induced bends and thereby causes a reduction in the conformational entropy of the polymer.[47]

The elastic modulus of a network of crosslinked semiflexible polymers depends on network connectivity.[3] When the network is well-connected, it deforms in an affine (i.e. uniform) manner. In this case, all filaments experience the same deformation and are predominantly stretched. The network elasticity can then be calculated analytically from an orientational average over the force–extension response of each filament.[29,47] In contrast, when network connectivity is low, it can be more energetically favorable for the filaments to bend, rather than stretch, in response to an applied shear stress, resulting in nonaffine deformations.[48-52] An analytical prediction of the elastic modulus is challenging in this case, and thus the nonaffine regime has mostly been explored by computer simulations.

Here we will compare our experimental data to analytical predictions assuming an affine network response. If nonaffinity is present, we expect it generically to decrease network stiffness compared to the affine limit, since nonaffinity increases the number of degrees of freedom in the system. In the affine limit, the network elastic modulus in the linear elastic regime, $G_0$, can be expressed in terms of the total fiber length per unit volume, $\rho$, and two length scales, namely $l_p$ and the distance between crosslinks, $l_c$:[47]

$$G_0 = 6 \rho k_B T l_p^2 / l_c^3. \quad (1)$$

The crosslink distance can be estimated using scaling theories for semiflexible polymers. Crosslinking is expected to occur either at the scale of the mesh size, $\xi \propto \rho^{-1/2}$, or at the scale of the somewhat larger entanglement length, $l_e$, which scales as $l_p^{1/5} \rho^{-2/5}$.[47,53,54] Once the shear stress exceeds a critical value, networks of semiflexible polymers will strain-stiffen as a consequence of the entropic resistance of the filaments to stretching. The shear stress characterizing the onset of nonlinearity, $\sigma_0$, can be expressed as:[47]

$$\sigma_0 = \rho k_B T l_p / l_c^2. \quad (2)$$

For fibers with a stretch modulus $\kappa_s$, the elastic modulus is expected to saturate at a plateau value $K_s = f \rho \kappa_s$, where $f$ is a geometrical prefactor that lies between $f = 1/15$ in the limit of isotropic network and $f = 1/8$ for highly aligned network.[9,55] Full expressions for the stress-dependent modulus can be calculated numerically.[29]

When the filaments comprising the network are themselves bundles of semiflexible polymers, the elasticity of the network becomes a function of the degree of bundling. In the case of fibrin, the fibers are bundles of $N_p$ protofibrils. Henceforth we take the superscript '$F$' to denote bundles of protofibrils (i.e. fibers) and '$pf$' to denote single protofibrils. Thus $\rho^{pf}$ is the length density of protofibrils, and the corresponding length density of fibers can be expressed as: $\rho_F = \rho^{pf} / N_p$. The persistence length of a bundle of wormlike chains can be predicted from the number of constituent chains and the effectiveness of cross-links in mechanically coupling adjacent filaments.[13] In particular, the bundle stiffness is bounded by two limits. In the fully coupled bending limit, which occurs when the shear stiffness of the crosslinks is large, the bundle behaves like a homogeneous elastic beam with $\kappa^F = N_p^2 \kappa^{pf}$. In the fully decoupled bending limit, which occurs when the cross-link shear stiffness is small such that protofibrils can easily slide relative to each other, the bending stiffness is linear in $N_p$ according to $\kappa^F = N_p \kappa^{pf}$. These behaviors can be summarized by the simplified expression,

$$l_p^F = N_p^x l_p^{pf}, \quad (3)$$

where $x$ is a coupling exponent that describes the strength of the linkage between the protofibrils in a bundle.[5] Note that this expression neglects any length-scale dependence of the bundle stiffness.[13] The coupling exponent can range from 1, corresponding to loose coupling, to 2, corresponding to tight coupling. Combining eq. (1) with eq. (3), and assuming $l_c \sim l_e$, we obtain for the plateau modulus of a network of bundles:

$$G_0 \sim 6 \rho k_B T (l_p^{pf})^{7/5} (\rho^F)^{11/5} N_p^{7x/5}. \tag{4}$$

We will show that the coupling exponent $x$ for networks of fibrin bundles can be directly calculated from measured $G_0$ values using eq. (4) with the bundle size $N_p$ and protofibril persistence length $l_p^{pf}$ as inputs.

## RESULTS

### Varying bundle size

Under physiological conditions, fibrinogen self-assembles in a hierarchical manner, first forming double-stranded protofibrils, which then bundle into thicker fibers. Confocal microscopy of these so-called '*coarse*' fibrin networks show an open meshwork made up of thick fibers, with a pore size of several microns (Fig. 1A). Transmission electron microscopy (TEM) images of fibrin networks deposited and dried on grids reveal that the fibers have a diameter in the range of 50–100 nm (Fig. 1C and Fig. S1A). Similar results were obtained by scanning electron microscopy (SEM) of fixed 3D fibrin networks (Fig. S2).

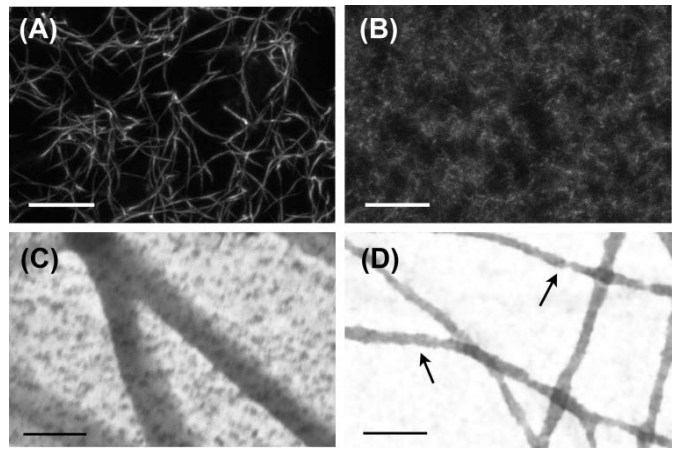

Fig. 1 Microstructure of (A,C) networks of bundled protofibrils under near-physiological ("*coarse* fibrin") conditions, and (B,D) networks of protofibrils that are barely bundled, prepared under high salt and high pH ("*fine* fibrin") conditions. Panels (A) and (B) show confocal fluorescence microscopy images of 1 mg/ml fibrin networks (scale bars 10 μm). Panels (C) and (D) show TEM images, revealing the fiber diameters (scale bars 200 nm). The arrows indicate evidence of protofibril twisting.

The goal of this work is to elucidate the contribution of each hierarchical structural level to the mechanical properties of fibrin networks. Even though fibrin fibers cannot be taken apart into their constituent protofibrils and linkers, since bundling is an intrinsic property of the protofibrils, we can exploit the known sensitivity of fibrin polymerization to salt and pH to vary the bundle size.[44-46] We assembled fibrin networks under conditions where lateral assembly of protofibrils is almost completely inhibited. This so-called '*fine*' fibrin limit is favored in a buffer with high pH (8.5) and high ionic strength (0.45). Confocal imaging reveals that *fine* fibrin indeed form a dense network with a pore size too small to visualize by optical microscopy (Fig. 1B). This is consistent with estimates of the pore size based on protofibril length density according to $\xi = \rho^{-1/2}$, which predicts a mesh size of ~150 nm at 1 mg/ml. TEM and SEM images confirm that the *fine* fibrin networks are composed of thin filaments (Fig. 1D and Fig. S2). The diameters of the *fine* fibers as determined from TEM images range from 15 to 30 nm (Fig. S1B). Structural models based on prior EM[45,56,57] and AFM[58,59] imaging indicate a diameter of single protofibrils in the range of 10–20 nm. Thus, the TEM data suggest that *fine* fibrin networks contain single protofibrils as well as bundles of 2 or maybe 3 protofibrils. However, the diameter measurements by themselves are not entirely conclusive, since drying, surface immobilization and observation in vacuum may influence the apparent diameter. Nevertheless, some fibers in the TEM images show clear evidence of $N_p = 2$, since fiber twisting can be distinguished.

Given the uncertainties involved with EM analysis of fiber diameters, we also measured $N_p$ based on wavelength-dependent light scattering (*turbidimetry*; see Experimental section) from fibrin networks in their hydrated state. For *fine* fibrin networks, we find an average $N_p$ value close to 2, independent of protein concentration, $c_p$ (Fig. S3). This observation confirms that minimal protofibril bundling occurs under *fine* fibrin conditions. By contrast, turbidimetry reveals that the average bundle size in *coarse* fibrin networks prepared under near-physiological conditions is close to 87 when the networks are formed at $c_p$ between 0.1–3 mg/ml, and thereafter decreases to reach a value of 20 at 8 mg/ml (Fig. S3).

**Rheology of *fine* fibrin networks**

To enable a quantitative interpretation of the mechanics of physiological (*coarse*) fibrin networks, which consist of protofibril bundles, we first study *fine* fibrin networks, which show minimal bundling. We probed the nonlinear elastic response of the networks by applying a stepwise increasing constant shear stress while superposing a small oscillatory stress to probe the tangent elastic modulus, $K'$. All networks strongly stress-stiffen, as shown in Fig. 2A. Depending on concentration, the networks can stiffen up to 100-fold before they break. The corresponding strain at rupture approaches values close to 200%, in line with the known elastomeric properties of fibrin.[41] The linear modulus measured at small strains, $G_0$, increases strongly when $c_p$ is raised from 0.5 to 6 mg/ml. More specifically, $G_0$ increases as a power law in $c_p$ with an exponent of 2.1±0.1 (solid squares in Fig. 3A). This exponent is consistent with the analytical model for networks of semiflexible polymers (Eq. (4)), which predicts an exponent of 11/5 (solid line in Fig. 3A). For reference, we note that our data agree well with prior measurements on fibrin networks prepared under similar *fine* fibrin conditions (Fig. S4A).[46,60]

Past a certain critical stress $\sigma_0$, $K'$ increases with stress in a complex fashion. Entropic models of crosslinked networks of inextensible semiflexible polymers such as actin[61] and intermediate filaments[53] predict a strong increase in stiffness with stress according to $K' \propto \sigma^{3/2}$.[47] We find that *fine* fibrin networks show a significantly weaker stress-dependence (Fig. 2B). Moreover, the extent of stiffening is dependent on $c_p$. A similarly weak stiffening with stress was previously seen for fish fibrin[29] as well as for vimentin.[53] In those studies, the weak stiffening response was attributed to filament backbone stretching, which is an enthalpic effect.[29,53] To test whether backbone stretching can also account for the stress response of the *fine* fibrin networks, we fitted the stiffening curves to the full theoretical prediction for the stiffness of networks of extensible semiflexible polymers.[29] This fitting requires 3 fit parameters: $l_p$, $\kappa_s$ and $l_c$. As shown in Fig. 2A, the model (solid lines) is indeed able to capture both the onset of strain-stiffening, which originates from chain entropy, and the inflection at intermediate stress, which stems from backbone stretching. However, at large stress, the model systematically underestimates the measured $K'$. The model accounts for shear-induced alignment of fibrin protofibrils under stress, but assumes that the stretch modulus of the protofibrils is independent of strain. The systematic discrepancy between the data and the predictions thus strongly suggests that the protofibrils themselves stiffen under extension, while the close agreement at low and intermediate stresses indicates that the minimally bundled protofibril networks are well represented as networks of extensible semiflexible polymers.

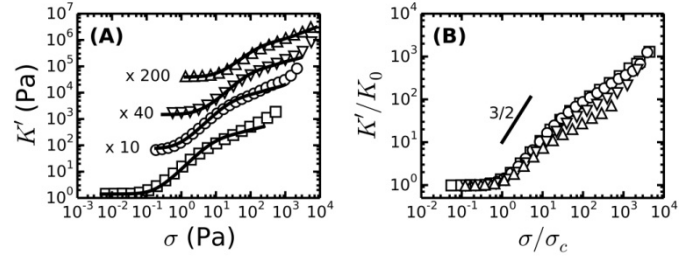

**Fig. 2** Stress-stiffening response of fibrin networks prepared under *fine* network condition near the single protofibril limit ($N_p = 2$), to an applied shear stress at different protein concentrations: $c_p = 0.5$ (squares), 1 (circles), 3 (triangles up) and 6 mg/ml (triangles down). In (A), the solid lines represent predictions of the affine thermal model for extensible chains. The $K'$ values have been shifted vertically for clarity, as indicated. (B) Normalized stress-stiffening curves. *Fine* fibrin networks show clear evidence of fiber stretching at high strain, since the stiffening curve is weaker than the predicted 3/2-scaling (short solid line) expected for inextensible polymers.

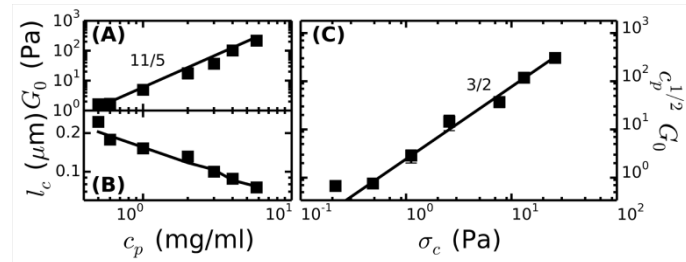

**Fig 3** Comparison of rheology of *fine* fibrin networks (squares) with the affine thermal model for extensible wormlike chains (lines). (a) Plateau modulus, $G_0$, as a function of fibrin concentration, $c_p$, compared with the predicted 11/5 power-law. (b) Cross-link distance, $l_c$, as a function of $c_p$ compared with the theoretical prediction for $l_c = l_e$. (c) $c_p^{1/2} G_0$ shows a power-law dependence on $\sigma_0$ with an exponent of 3/2, consistent with the notion that $G_0$ and $\sigma_0$ are both governed by the entropic force–extension behavior of the filaments.

Let us examine whether the obtained fit parameters, $l_p$, $\kappa_s$ and $l_c$, are physically meaningful. We obtain a value of 150 nm for the average $l_p$ of the filaments forming the *fine* fibrin networks, independent of $c_p$ (Fig. S5). Given an average $N_p$ of 2, this corresponds to $l_p^{pf} = 75$ nm, consistent with the supposition that the protofibrils are semiflexible polymers. This value is smaller than estimates from light scattering experiments (200 nm)[62] and analysis of EM images of fish fibrin (500 nm),[29] but is close to values reported in recent light scattering and small-angle X-ray scattering experiments (120 nm).[63] The apparent $\kappa_s^{pf}$ obtained from the fits (assuming $N_p = 2$) lies between 80 and 150 pN (closed symbols in Fig. S6A), close to the range (50–100 pN) inferred from macroscopic rheology measurements on fish fibrin.[29] If we assume that the protofibrils behave as homogeneously elastic cylinders of diameter 10 nm, we can infer from the $\kappa_s^{pf}$ values a Young's modulus $E$ between 1 and 1.9 MPa, which is at the low endof the range of 1.7–15 MPa measured by bending and stretching of fibrin fibers.[40,64,65] Fibrin protofibrils are therefore somewhat softer than intermediate filaments ($E = 9$ MPa)[53] and three orders of magnitude softer than actin filaments ($E = 1$–3 GPa).[66] The apparent $l_c$ obtained from the fits decreases from 0.25 μm at 0.5 mg/ml (1.5 μM) fibrin to 0.05 μm at 6 mg/ml (17 μM) fibrin (symbols in Fig. 3B). Remarkably, these fitted values of $l_c$ closely agree with the values of $l_e \sim l_p^{1/5} \rho^{-2/5}$ predicted by scaling theory[47,53,54] if we assume a prefactor of 0.75 (Fig. 3B, solid line), and they are close to the lower bound expected in case of dense crosslinking, $l_c \sim \rho^{-1/2}$ using the same prefactor (Fig. S4B).

As a further test of the applicability of the entropic model for extensible semiflexible polymers to *fine* fibrin networks, we examined the relationship between $G_0$ and the onset of stiffening, $\sigma_0$. Combining Eq. (1) and Eq. (2), the model predicts that $c_p^{1/2} G_0 \propto \sigma_0^{3/2}$. Our data are indeed consistent with this prediction (Fig. 3C). Together with the fact that we obtain physically meaningful values of $l_p$, $\kappa_s$ and $l_c$, this strongly supports our conclusion that entropic elasticity combined with a transition to enthalpic elasticity once the filaments are pulled out underlies the nonlinear elastic response of *fine* fibrin networks.

**Bundled networks: *coarse* fibrin networks**

Under physiological conditions of blood clotting, fibrin protofibrils laterally aggregate to form bundles.[34] To mimic these conditions, we formed fibrin networks at near-physiological pH (7.4), ionic strength (0.17 mM) and temperature (37°C). We previously observed that the stiffness of such '*coarse*' networks increases with $c_p$ with an exponent close to 11/5,[9] similar to the trend in *fine* fibrin networks, and consistent with the predicted exponent for semiflexible polymers (Fig. 4A). We thus hypothesize that the networks are composed of fibers that can be modeled as semiflexible bundles of protofibrils. To test this hypothesis, we will now compare the rheology of bundled protofibril networks to the *fine* fibrin limit. In particular, Eq. (4) allows us to calculate $x$, characterizing the tightness of protofibril bundles in *coarse* fibrin networks, by using the measured $G_0$ as input from rheology (Fig. 4A) and the measured $N_p$ from turbidimetry (Fig. 4B). Using $l_p^{pf} = 75$ nm, as determined from the analysis of *fine* fibrin rheology, we find values for $x$ close to 2, demonstrating that fibrin fibers behave as tight bundles of protofibrils (Fig. 4C).

We further test whether the rheology of *fine* and *coarse* fibrin networks can be reconciled by the different degrees of bundling. From Eq. (4), we expect $G_0 \propto (\rho^F)^{11/5} (l_p^{pf})^{7/5} N_p^{7x/5}$. As shown in Fig. 5A, the *fine* and *coarse* fibrin data sets are indeed entirely consistent, using $x = 2$ for *coarse* and $x = 1$ for *fine* fibrin. Moreover, both data sets agree well with the theoretical model assuming an affine network deformation (solid line). The model (Eqs. (2) and (3)) also predicts $\sigma_0$ to scale as $(\rho^F)^{9/5} (l_p^{pf})^{3/5} N_p^{3x/5}$. Again, we found good agreement of *coarse* as well as *fine* fibrin data with the affine model (Fig. 5B). This agreement also shows that the *coarse* fibrin networks deform rather affinely. To test whether this rescaling holds over an even wider range of bundle sizes, we prepared networks with bundles of ~366 protofibrils by removing fibrinogen oligomers prior to polymerization by gel filtration.[67] As shown by the gray circle in Fig. 5, $G_0$ and $\sigma_0$ of this highly bundled network are also consistent with the prediction of the semiflexible bundle model, but with a smaller coupling exponent of $x = 1.3$.

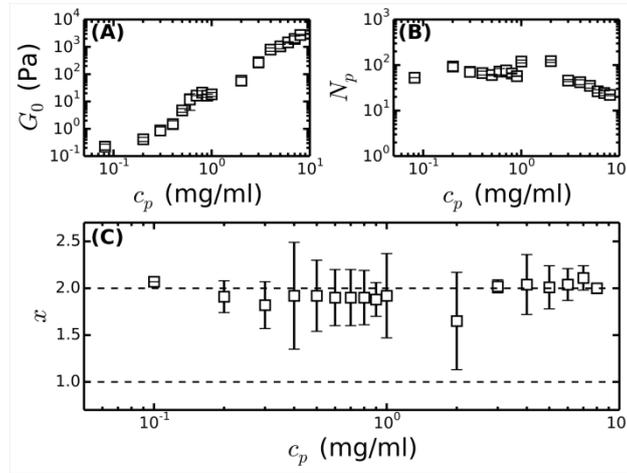

**Fig. 4** Bundle coupling in *coarse* fibrin networks formed under near-physiological conditions, inferred by comparing the linear elastic modulus of bundled (*coarse*) fibrin networks to that of *fine* networks prepared close to the unbundled protofibril limit. (A) Measured plateau modulus ($G_0$) of *coarse* fibrin, taken from our previous work.[9] (B) Bundle size ($N_p$) of these same networks, obtained by reanalyzing our previous turbidity data[9] with a recently proposed, more accurate scattering model.[23] (C) Bundle coupling strength expressed in terms of the exponent $x$, which is calculated using eq. (4), based on the data in (A) and (B), and assuming $l_p^{pf} = 75$ nm. The horizontal dashed lines indicate the two limits for $x$, where $x = 1$ corresponds to loose coupling and $x = 2$ corresponds to tight coupling.

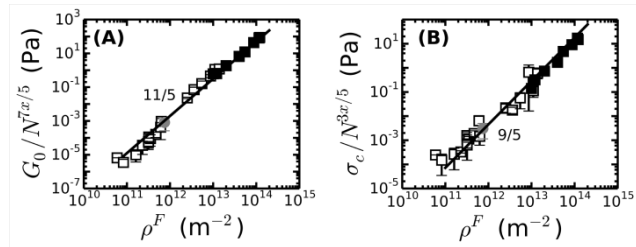

**Fig. 5** The concentration dependences of (A) the plateau modulus $G_0$ and (B) the onset stress for stress-stiffening $\sigma_0$ of fibrin networks are both consistent with the semiflexible bundle model. This is demonstrated by rescaling $G_0$ and $\sigma_0$ for *fine* (solid squares) and *coarse* fibrin networks (open squares) by their dependence on $N_p$. The gray circle represents a fibrin network of 1 mg/ml prepared from gel-filtered fibrinogen, for which $N_p = 366$. Coupling is assumed to be loose in case of *fine* fibrin ($x = 1$) and tight for *coarse* fibrin ($x = 2$, see Fig. 4), and is found to be intermediate for the ultra-thick fibrin fibers ($N_p = 366$, $x = 1.3$).

We next compared the stress-stiffening behavior of *coarse* and *fine* fibrin networks beyond $\sigma_0$, where network elasticity is dominated by enthalpic stretching of the polymer backbones. The simplest hypothesis is that the protofibrils will stretch in parallel, which we can test by rescaling $K'$ and $\sigma$ by the length density of protofibrils, $\rho_{pf}$. As shown in Fig. 6, this normalization collapses the stiffening curves of *coarse* and *fine* fibrin networks onto a single master curve once the average force per protofibril reaches ~1 pN. Importantly, this universal response is independent of $N_p$, $x$ or $c_p$, indicating that in the nonlinear regime, the protofibrils contribute independently to the network elasticity. Thus the (enthalpic) stretch modulus of fibrin fibers is linear in the number of constituent protofibrils. This behavior may also reflect a decoupling of the tight bundle structure that is theoretically expected for short-wavelength deformations:[13] Tight bundle behavior is expected for long wavelength bending, which dominates at low stress, while increasingly loose bundle behavior is expected for shorter wavelength bends that are dominant under high axial loads.

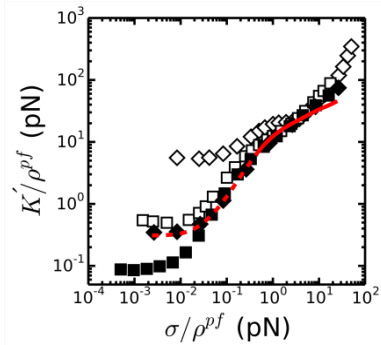

**Fig. 6** Direct comparison of the high-stress (enthalpic) elastic response of bundled fibrin networks prepared under *coarse* conditions (open symbols) and *fine* fibrin networks (closed symbols). Data are shown for two protein concentrations: $c_p = 0.5$ (squares) and 3 (diamonds) mg/ml. The affine thermal model prediction for extensible wormlike chains for 3 mg/ml *fine* fibrin (red dashed line) is shown, where the enthalpic regime is indicated by the solid line.

The normalized stiffening curves above $\sigma/\rho^{pf} \sim 1$ pN show two regimes. The stiffening first shows an inflection, which is visibly tending to a plateau for *coarse* fibrin networks. This plateau region is presumably governed by the linear (small-strain) axial stretch modulus of the protofibrils. When the force reaches $\sigma/\rho^{pf} \sim 5$ pN, the stiffening enters a second regime where $K'/\rho^{pf}$ starts to increase again. The apparent $\kappa_{s,}^{pf}$ in the linear elastic stretch regime that we can thus estimate from the normalized data for *coarse* (bundled) fibrin is ~100–350 pN if we assume that the network is still isotropic network and ~60–200 pN if we assume that the network is aligned, consistent with the $\kappa_{s,}^{pf}$ that we estimated above from *fine* fibrin rheology (80–150 pN; see Fig. S6A). The better agreement with the assumption of aligned network is consistent with the fact that the strain level corresponding to the linear portion of the enthalpic regime falls around 30% strain (Fig. S6B), where we already expect significant fiber alignment.

If protofibril stretching were strictly linear, only a weak increase of the network stiffness would be expected at high strain, reflecting further shear-induced fiber alignment. Strikingly, the model prediction in Fig. 6 (solid red line) systematically underestimates the actual stiffness of both *fine* and *coarse* fibrin networks at large stress. This discrepancy suggests that fibrin protofibrils intrinsically stiffen beyond a certain level of stretch. This hypothesis is indeed supported by prior force–extension measurements by AFM on individual fibers, which showed strain-stiffening of the fibers.[39,40,65]

**Varying bundle tightness**

We have shown evidence that fibrin reconstituted under near-physiological conditions can be modeled as a network composed of tightly coupled protofibrils over a range of bundle sizes. The stiffness of a wormlike bundle is expected to be strongly dependent on the coupling strength between the constituent polymers. Previous studies have shown that lateral association of protofibrils is promoted by crosslinking of long, flexible α-chains protruding from the protofibril surfaces, as sketched in Fig. 7A.[34,68,69] Crosslinking is mediated by the enzyme FXIII, which creates covalent peptide bonds between specific sites on the α-chains. FXIII additionally creates crosslinks between α- and γ-chains, as well as crosslinks between γ-chains within protofibrils.[70,71] Based on this evidence, we hypothesize that FXIII-mediated crosslinking may control the tightness of the bundle.

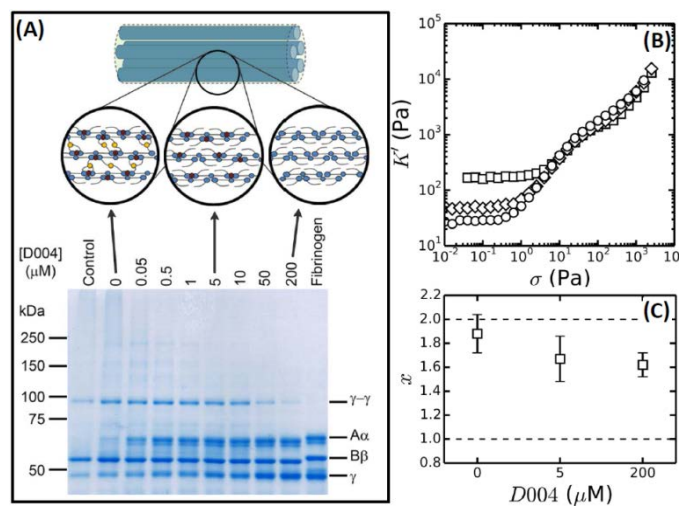

**Fig. 7** Influence of FXIII-mediated crosslinking on fibrin bundle stiffness. (A) Reducing SDS-PAGE gel of 2 mg/ml *coarse* fibrin networks formed in the presence of different concentrations of the FXIII-inhibitor D004. The sample marked "control" consists of a 2 mg/ml fibrin network without DMSO, showing that the addition of DMSO alone (without D004) results in slight reduction in the extent of α- and γ-chain crosslinking. The schematic on top depicts crosslinks between protofibrils (α–α-crosslinks) in yellow and intra-protofibril crosslinks (γ–γ-crosslinks) in red. (B) Stress-stiffening curves for 2 mg/ml fibrin networks with 0 μM (squares), 5 μM (diamonds) or 200 μM (circles) D004. (C) Corresponding coupling factor $x$ calculated from eq. (4). Crosslink inhibition makes the bundles less tight. The two limits ($x = 2$ for a tight bundle and $x = 1$ for a loose bundle) are indicated by horizontal dashed lines.

To test this hypothesis, we controlled the crosslinking activity of FXIII by adding an FXIII inhibitor, D004, in the range of 0 μM to 200 μM.[38] As shown by SDS-PAGE analysis of solubilized and denatured fibrin networks, we can inhibit crosslinking in a graded manner (Fig. 7A). At 0 μM D004, there is both α-crosslinking (distinguishable by the disappearance of the band corresponding to the monomeric α-chain) and γ–γ-crosslinking (bands indicated). When we add 5 μM D004, there is no detectable α-chain crosslinking (neither $α_N$ polymers nor $α_N–γ_M$ crosslinks), consistent with prior reports,[38,70] and the amount of γ–γ-crosslinking is reduced to about 50%. There is complete inhibition of crosslinking at 200 μM D004 (see Fig. S8A for the quantification).

To test how crosslinks influence bundle rigidity, we measured stress-stiffening curves of fibrin gels upon FXIII inhibition, where we selected 0, 5 and 200 μM D004 (respectively squares, diamonds and circles in Fig. 7B). Complete inhibition of α-crosslinking at 5 μM D004 causes a drop in linear elastic modulus by a factor of 4. Full inhibition of FXIII further reduces the linear modulus by a factor 6 compared to the control, consistent with previous reports.[38,70,72,73] According to the proposed semiflexible bundle model, the small-strain regime is determined by entropic elasticity, and should thus be sensitive to the bending rigidity of the fibers. A smaller $G_0$ indicates a looser, more flexible bundle, assuming that $N_p$ stays constant. Using eq. (4), we calculated the bundle coupling exponent, $x$, directly from the rheology data taking into account the slight change of $N_p$ with D004 level measured by turbidimetry (Fig. 8B). As shown in Fig. 7C, $x$ decreases from a value close to 2 for fully crosslinked networks to 1.6 for uncrosslinked networks. Interestingly, $x$ is still significantly larger than 1 in the absence of crosslinking. This means that, even in the absence of covalent crosslinks, protofibril bundles are still rather tightly coupled. Similarly, a decrease in gel stiffness upon reduced internal fiber crosslinking has been observed for other semiflexible bundle systems, such as actin bundled with fascin[18] and nanotubes bundled via covalent crosslinks.[14]

Strikingly, after the onset of strain-stiffening, the stress-stiffening curves overlap for all three D004 concentrations (Fig. 7B). This observation is consistent with the data presented in Fig. 6, which likewise show that the high-strain regime is determined by independent stretching of the protofibrils. To test whether inhibition of FXIII crosslinking has any effect on the stretch modulus of the protofibrils through changes in γ–γ-crosslinks, we also performed rheological measurements on *fine* fibrin network with varying levels of D004. SDS-PAGE revealed a gradual decrease of γ–γ- and α-chain crosslinking with D004, with lower concentrations of D004 required to inhibit crosslinking of *fine* fibrin compared to *coarse* fibrin (Figs. S8B and S9). Inhibition of α-chain crosslinking was complete at 0.05 μM D004, while γ–γ crosslinking was inhibited completely at 1 μM. Strikingly, crosslink inhibition by addition of D004 does not change $\kappa_s^{pf}$, even at concentrations where γ–γ-crosslinking is completely inhibited (Fig. 8A). We conclude that crosslinking of protofibrils within protofibril bundles increases the linear elastic modulus of *coarse* networks by increasing the fiber bending rigidity, but does not change the enthalpic elastic response of the networks at high stress.

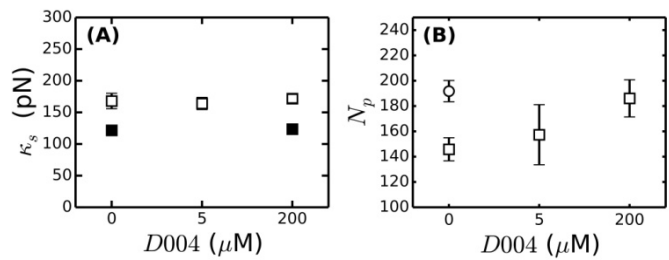

**Fig. 8** (A) The stretch modulus of fibrin protofibrils does not change when crosslinking by FXIII is inhibited by D004, assuming aligned networks in the enthalpic stretch regime. All networks are with 1% final DMSO concentration. (B) Influence of crosslink inhibition by D004 on in the bundle size $N_p$ of *coarse* fibrin based on turbidity measurements. Open squares are for 2 mg/ml *coarse* fibrin with 1% DMSO final concentration, while the open circle represent the 0% DMSO control, while closed squares are for *fine* fibrin.

## DISCUSSION

Polymer bundles are present in many biological systems, from cytoskeletal components inside the cell to extracellular matrix in tissues. The cross-sectional size of these bundles can vary from a few monomers in case of cytoskeletal bundles to hundreds or thousands in case of extracellular matrix fibers. In the case of fibrin networks, the size of the fibers can be tuned by changing pH and salt conditions. Here we have shown how we can dissect the properties of these bundles of semiflexible polymers by comparing the mechanical properties of fibrin networks prepared with different levels of bundling. We experimentally varied the average size of fibrin bundles from 2 up to 366 constituent protofibrils, thus changing bundle size by more than two orders of magnitude. We demonstrated that the nonlinear rheology of networks close to the limit of unbundled protofibrils is in excellent quantitative agreement with theoretical predictions for networks of semiflexible polymers.

By comparing the rheology of *coarse* and *fine* fibrin networks, we demonstrated that the fibers in *coarse* networks behave mechanically as bundles of protofibrils. By comparing the linear elastic modulus of *coarse* and *fine* networks we could directly quantify the coupling strength, $x$, which is close to 2 when the network is fully crosslinked. In this limit, the fibers behave as tightly coupled bundles of protofibrils. Based on $l_p^{pf} = 75$ nm, we expect $l_p$ for fibers in *coarse* networks to range from 30 µm at 8 mg/ml (where $N_p \sim 20$) to 560 µm at concentrations below 3 mg/ml (where $N_p \sim 86$). We indeed showed in earlier work that fibrin fibers within *coarse* networks exhibit measurable thermal fluctuations that show up as a $\omega^{3/4}$ frequency spectrum in high-frequency optical tweezers microrheology.[9] We note that at the time we concluded that the bundles had to be loosely coupled to account for this semiflexibility, but this conclusion was based on published values for $l_p^{pf}$ of 500 nm.[29] We now find a significantly smaller $l_p^{pf}$ of only 75 nm by rheology measurements on *fine* fibrin, which can account for the semiflexible behavior we observed as the fibers are tightly coupled. This is furthermore consistent with recent estimation using light scattering and small-angle X-ray scattering experiments.[63] It will be interesting in the future to measure the persistence length of protofibrils directly by single-fibril stretching with AFM or optical tweezers.

For *coarse* networks made of fibers with a bundle size of 366, the coupling exponent was 1.3, closer to the limit of loose bundling. This loosening may potentially be explained by a change in molecular packing with increasing fiber diameter. The cross-sectional molecular packing structure of fibrin is still poorly understood. Turbidity studies have shown that typically less than 30% of the volume of the individual fibrin fibers is comprised of protein.[23] Small Angle X-ray Scattering experiments[74-76] and diffraction analysis of electron microscopy images[76,77] have variably indicated either disordered or partially ordered lateral order. Stretching experiments and fluorescence intensity measurements on single fibrin fibers suggested a fractal-like packing.[78] The packing structure and therefore the effective bundle coupling exponent may thus well be dependent on bundle size.

Crosslinking by FXIII acts to enhance bundle tightness. However, this only influences the linear elastic modulus of the network and the onset for strain-stiffening, which are both determined by the entropic elasticity of the bundles. By contrast, the network stiffness in the nonlinear regime is insensitive to crosslinking. In particular, when we rescale the nonlinear mechanics to the total protofibril length per volume, $\rho^{pf}$, the high-strain response of *fine* and *coarse* networks overlap. Thus, in the enthalpic elastic regime, the protofibrils are simply stretched in parallel. Rheology measurements on *fine* fibrin of varying degrees of crosslinking also reveal that the stretch modulus of the individual protofibrils is not affected by crosslinking.

The hierarchical structure of fibrin fibers results in a hierarchical mechanical response, with an entropic linear elastic regime, followed by entropic stiffening, then enthalpic stretching and finally stretch-stiffening of the fibers themselves when the average force per protofibril exceeds ~10 pN. At high stress, both *fine* and *coarse* fibrin networks stiffen more than predicted by the wormlike chain model that assumes a linear elastic response for the protofibrils (Fig. 6). This interpretation is consistent with direct force–extension measurements by AFM on individual fibers, which showed that fibers stiffen at tensile strains in excess of ~100%.[39,40,65]

Several different interpretations for intrinsic nonlinearity of fibrin fibers have been proposed. One interpretation is that the supramolecular structure of the fibers is responsible for fiber stiffening.[29,40] The protofibrils are coupled by long and rather flexible carboxy-terminal extensions of the Aα-chains (αC region) that protrude from the protofibrils.[69] The combination of flexible elements with more rigid folded elements may give rise to nonlinearities once the flexible elements are fully stretched.[40] Support for this idea comes from force–extension measurements on fibers assembled from fibrinogen of different species, which demonstrated that a longer Aα-chain length correlates with greater extensibility.[79] Moreover, molecular dynamics simulations also indicated that the αC regions can play a crucial role in fibrin fiber mechanics.[80] However, here we observe intrinsic nonlinearity also in *fine* fibrin, which is minimally bundled, and this nonlinearity contributes equally to *fine* and *coarse* network stiffening. This finding argues against a supramolecular origin of nonlinearity, and instead suggests that the nonlinearity is intrinsic to the molecular structure of the protofibrils themselves. A likely source of protofibril nonlinearity is forced monomer unfolding. Molecular simulations showed that different domains within fibrin monomers start to unfold at forces in the range of 75–150 pN, accompanied by a conversion of the α-helical coiled-coil connector regions into stiffer β-sheet structures.[81] It is *a priori* difficult to predict how this unfolding behavior will be modified once fibrin monomers are incorporated in the double-stranded structure of a protofibril or the even large structure of a fiber. However, Fourier transform infrared spectroscopy[82] and direct staining of stretched networks with β-sheet-specific dye Congo Red[83] showed convincing evidence of a strain-induced conversion of α-helical into β-sheet secondary structure. Single protein unfolding measurements indicate typical forces of 90 pN to unfold fibrin monomers,[84] which is comparable to the largest forces per monomer that can be applied during shear rheometry without network breakage (~100 pN, see Fig. 6). To directly resolve the microscopic origin of protofibril stiffening under shear, it will be important to perform *in situ* measurements of fibrin secondary structure in combination with shear rheometry using, for instance, vibrational spectroscopy or X-ray scattering techniques.

## CONCLUSION

Here we have shown that both small and large bundles of fibrin protofibrils give rise to a rich mechanical response to an applied shear stress. The fibers can be modeled as bundles of protofibrils, whose bending rigidity increases quadratically with bundle size whereas the stretch rigidity increases only linearly with bundle size. At high strain, the bundles exhibit elastomeric properties and strong strain-stiffening. Altogether, the entropic and enthalpic elasticity of fibrin fibers protect fibrin networks against mechanical deformations. Our findings have important implications for understanding the origins of fibrin mechanics, especially in the contexts of bleeding disorders associated with defective crosslinking[71] and thrombosis associated with excessive stiffness.[85] Furthermore, these results can inspire the design of new bioinspired hierarchical materials with tunable mechanical properties.

## EXPERIMENTAL

### Fibrin polymerization

To obtain fibrin networks close to the protofibril limit (traditionally referred to as '*fine* clots'),[86] human fibrinogen (FIB3, Enzyme Research Laboratories, Swansea, UK) was dialyzed for 2 days at 4°C against a 50 mM Tris-HCl/400 mM NaCl buffer with an ionic strength 0.45, as described previously.[44-46] The pH was adjusted to 8.5 with NaOH. The dialyzed fibrinogen was centrifuged for 20 minutes at 9000 rpm to remove any aggregates. The final protein concentration was determined by spectrophotometrically by determining the absorbance at a wavelength of 280 nm with correction for scattering at 320 nm.[46] *Fine* fibrin networks were polymerized by adding 0.5 U/ml human thrombin (Enzyme Research Laboratories) in *fine* fibrin buffer (50 mM Tris-HCl/400 mM NaCl, pH 8.5) in the presence of 3.2 mM $CaCl_2$ at 37°C.

Data from *fine* fibrin were compared to data for networks of bundled protofibrils, often referred to as '*coarse* clots'. FIB3 fibrinogen was diluted in a buffer of near-physiological pH and ionic strength (20 mM HEPES/150 mM NaCl, 5 mM $CaCl_2$, pH 7.4). Polymerization was initiated by adding 0.5 U/ml thrombin and incubating the samples at 37°C. As the fibrinogen stock solution contains FXIII, the fibrin networks contained a constant molar ratio of FXIII to fibrinogen at all fibrinogen concentrations. The networks were always fully crosslinked, as shown by sodium dodecyl sulfate polyacrylamide gel electrophoresis (SDS-PAGE) analysis (Fig. S10). Data for crosslinked *coarse* fibrin networks were taken from our own earlier work.[9] New data for *coarse* fibrin with reduced levels of crosslinking were obtained by adding a specific FXIII inhibitor, 1,3-Dimethyl-4,5-diphenyl-2-[(2-oxopropyl)thio]imidazolium trifluorosulfonic acid salt (D004)[87] before thrombin addition. D004 was obtained from Zedira (Darmstadt, Germany) and dissolved in dimethylsulfoxide (DMSO) to a concentration of 20 mM. We used D004 concentrations between 0 and 200 μM. Since DMSO can affect fibrin assembly,[88] we used a constant DMSO concentration of 1%v/v for all tests, including controls, involving FXIII inhibition.

We furthermore obtained new data for *coarse* networks of fibrin fibers with exaggerated bundling (on average 366 protofibrils per bundle, compared to ~90 for the standard *coarse* fibrin). These networks were obtained by using gel filtration to remove oligomers from the FIB3 fibrinogen stock.[67] Briefly, FIB3 fibrinogen was filtered through a 0.2 μm filter and injected at a concentration of 2.7 mg/ml onto a Superdex 200 column that had been equilibrated with fibrin buffer (20 mM HEPES/150 mM NaCl, pH 7.4). The flow rate was 0.5 ml/min at a pressure of ~0.12 MPa and at room temperature. The chromatograms showed two peaks, with the first peak corresponding to fibrinogen oligomers and the second peak to fibrinogen monomers. The monomer fraction was concentrated to ~15 mg/ml using MacroSep centrifuge tubes (Pall Corporation) at 811 rcf. The tubes were washed with buffer before use. The final protein concentration was again determined by spectrophotometry. The fibrinogen monomer stock was snap-frozen and stored at –80°C. SDS-PAGE analysis showed that FXIII was still present in the preparation since both α-polymers and γ–γ-dimers were seen on the gel.

**Rheology**

The nonlinear viscoelastic properties of the fibrin networks were measured using a stress-controlled rheometer (Physica MCR 501; Anton Paar, Graz, Austria). Directly after thrombin addition, the fibrinogen solutions were quickly transferred to the rheometer, which was equipped with a steel cone and plate geometry (20, 30 or 40 mm diameter, 1° cone angle). The rheometer was preheated to 37°C. Solvent evaporation was prevented by coating the sample edges with mineral oil. The time evolution of the linear complex shear modulus, $G^*$, was monitored during fibrin polymerization by applying a small-amplitude oscillatory strain with amplitude $\gamma = 0.5\%$ and frequency $\omega = 3.14$ rad/s and by measuring the stress response, $\sigma(\omega) = G^* \gamma(\omega)$. The shear modulus is a complex quantity, $G^* = G' + iG''$, having an in-phase elastic component, $G'$, and an out-of-phase viscous component, $G''$. Networks of *fine* fibrin reached a constant shear modulus $G_0$ after about 1 hour, while *coarse* fibrin reached a steady state only after 4 hours.

To probe the nonlinear mechanical response, we used a differential measurement protocol, which captures the stress-stiffening response of biopolymer networks more accurately than large amplitude oscillatory shear measurements.[89] Briefly, small amplitude stress oscillations of amplitude $\delta\sigma = 0.1\sigma_0$ and frequency 0.1 Hz are superimposed on a steady shear stress, $\sigma_0$, that is gradually increased in a stepwise manner. The tangent shear modulus, which is the local tangent of the stress-strain curve, follows from the oscillatory strain response, $K^*(\sigma_0) = \delta\sigma/\delta\gamma$. $K^*$ has an in-phase elastic component, $K'$, and an out-of-phase viscous component, $K''$. In the linear response regime, $K'$ equals the linear elastic plateau modulus, $G_0$. The networks were nearly perfectly elastic and did not exhibit any significant creep until the shear stress was close to the breakage point. Moreover, the stiffening curves were repeatable as long as the stress did not exceed the rupture stress. Unless noted otherwise, the rheology data represent the mean ± standard deviation from at least three independent experiments.

**Imaging**

To measure the diameter of the fibers, we performed TEM using a Verios electron microscope (FEI Europe BV, Eindhoven, the Netherlands) operating at 20 kV. About 20 μl of freshly prepared fibrinogen–thrombin solution was quickly deposited as a thin layer on EM grids (Ted Pella, Van Loenen Instruments, Zaandam, the Netherlands) and polymerized at 37°C in a humid atmosphere. After complete polymerization (1 hour for *fine* fibrin, 4 hours for *coarse* fibrin), the grids were washed 5× with MilliQ water and air-dried. Samples were imaged the same day. Fiber diameters were measured manually. We counted more than 200 fibers, combining data from more than five randomly chosen fields-of-view of networks polymerized at concentrations between 0.5 and 2 mg/ml.

Scanning Electron Microscopy (SEM) was performed on fibrin samples using Verios electron microscope. Fibrin networks were polymerized in 20 μl dialyzing buttons (Hampton Research, Aliso Viejo, United States) in a humid atmosphere. After polymerization, the gels were washed 3x by cacodylate buffer (50 mM sodium cacodylate, 150 mM natrium chloride, pH 7.4), followed by 2 hours or overnight fixation with 2% glutaraldehyde in cacodylate buffer. After fixation, samples were washed 3x with cacodylate buffer and then dehydrated by increasing percentages of ethanol. After complete dehydration (100% ethanol), samples were washed with 50% hexamethyldisilazane (HMDS) in ethanol and twice with 100% HMDS. Samples were left overnight to evaporate residual HMDS under the hood. After complete HMDS evaporation, samples were transferred to stubs equipped with carbon tape and sputter coated with a 15.4 nm gold-palladium layer. Samples were imaged at 10 kV using secondary electrons.

To visualize the architecture of fibrin networks in their native, hydrated state, we performed confocal fluorescence microscopy using a Nikon Eclipse Ti inverted microscope equipped with a 100x oil immersion objective (NA 1.49), a 488-nm laser (Coherent, Utrecht, The Netherlands) for illumination, and a photomultiplier tube detector (A1; Nikon, Amsterdam, the Netherlands). AlexaFluor488-labelled fibrinogen was purchased from Life Technologies (Bleiswijk, the Netherlands), dissolved in either *fine* fibrin buffer or *coarse* fibrin buffer (without $CaCl_2$) and mixed with unlabeled fibrinogen in a 1:10 molar ratio. Samples were prepared in sealed glass chambers with a height of 0.25 mm and polymerized at 37°C for 1 hour (*fine* fibrin) or 4 hours (*coarse* fibrin) before imaging. The images shown are maximum intensity projection over stacks of 129 images over a total z-distance of 25.6 µm, taken 25 µm away from the bottom coverslip surface.

**Turbidity**

Since diameter estimates from TEM images are prone to artifacts from drying, surface attachment and observation in vacuum, we also measured the diameter and mass-length ratio, $\mu$, of fibrin fibers in their hydrated state by turbidimetry. These measurements were carried out using a Cary300 UV-Vis spectrophotometer (Agilent Technologies, Amstelveen, Netherlands). Fibrin gels were polymerized directly in disposable cuvettes (UV-Cuvette micro, Plastibrand, Germany), which were closed with caps to prevent solvent evaporation. To remove any air bubbles, cuvettes with 350 µl fibrinogen solution were degassed in vacuum for ~8 min, before starting polymerization at 37°C by the addition of thrombin.

Once the samples were fully polymerized, the optical density, *OD*, was measured as a function of wavelength, $\lambda$, between 350 and 900 nm. To extract the fiber dimensions from the turbidity, $\tau = OD \ln(10)$, we analyzed the data according to a theoretical model proposed by Carr et al[90] and later extended by Yeromonahos and co-workers.[23] Assuming that the networks can be modeled as isotropic networks of rigid cylindrical fibers with a large length-to-diameter ratio, the turbidity $\tau$ can be expressed in the form: $\tau\lambda^5 = A\mu (\lambda^2 - Ba^2)$. Here $\lambda$ is wavelength in cm, $\mu$ is the mass-length ratio in Da/cm, and $a$ is the fiber radius in cm. $A$ and $B$ are constants and are respectively equal to $(88/15)c\pi^3 n_s (dn/dc)^2 (1/N_A)$ and $(184/231) \pi^2 n_s^2$. Here, $N_A$ is Avogadro's number, $n_s$ is the refractive index of the solvent, $dn/dc$ is the specific refractive index increment ($dn/dc = 0.17594$ $cm^3 g^{-1}$ for fibrin [ref]), and $c$ is the fibrinogen concentration expressed in g/ml. Thus, $\tau\lambda^5$ is expected to be linear in $\lambda^2$ with a slope that is proportional to $\mu$ and a y-intercept that is related to both $\mu$ and $a$. Note that this expression includes a small correction of the original formulas in Ref. [23] (private communication, F. Caton). Given that individual protofibrils have a mass-length ratio $\mu_0 = 1.44 \times 10^{11}$ Da/cm,[91] the number of protofibrils in a fiber, $N_p$, is simply given by $N_p = \mu / \mu_0$. We observed a linear dependence of $\tau\lambda^5$ on $\lambda^2$ for both *coarse* and *fine* fibrin networks between 650 and 800 nm, and thus this range was chosen to fit the data. Turbidity data represent an average over three independent measurements per condition. Data for crosslinked *coarse* fibrin (i.e. without D004) were taken from a previous study,[9] but re-analyzed according to this corrected model. Since *fine* fibrin networks scatter rather weakly, we could only obtain reliable results for concentrations above 2 mg/ml, where the *OD* was above 0.01. In contrast, *coarse* fibrin networks scatter strongly, showing an *OD* above 0.05 at all concentrations tested.

**Crosslinking analysis by SDS-PAGE**

The extent of covalent crosslinking of the γ and α chains of the fibrin monomers incorporated into fibrin networks was analyzed by reducing SDS-PAGE analysis. *Coarse* fibrin networks over a range of concentrations (0.5–8 mg/ml), as well as *coarse* and *fine* fibrin networks in the presence of varying amounts of D004 (0–200 μM, 1% DMSO final concentration) at a fixed fibrin concentration of 2 mg/ml, were tested. Fully formed fibrin gels were dissolved by adding SDS-PAGE sample buffer (Sigma Aldrich, Zwijndrecht, the Netherlands) and heating at 95°C. Samples were run on 8% polyacrylamide gels, and stained with InstantBlue (Gentaur, Eersel, the Netherlands).

# ACKNOWLEDGEMENTS

The authors thank J. Weisel, R. Litvinov and C. Nagaswami for their protocol for performing SEM on fibrin clots and for helpful discussions. This work is part of the research programme of the Foundation for Fundamental Research on Matter (FOM), which is financially supported by the Netherlands Organisation for Scientific Research (NWO). This work is further supported by NanoNextNL, a micro- and nanotechnology programme of the Dutch Government and 130 partners. NAK acknowledges support by a Marie Curie IIF Fellowship.

# Hierarchical strain-stiffening of semiflexible wormlike bundles

Supplementary Information

**Supplementary Figures**

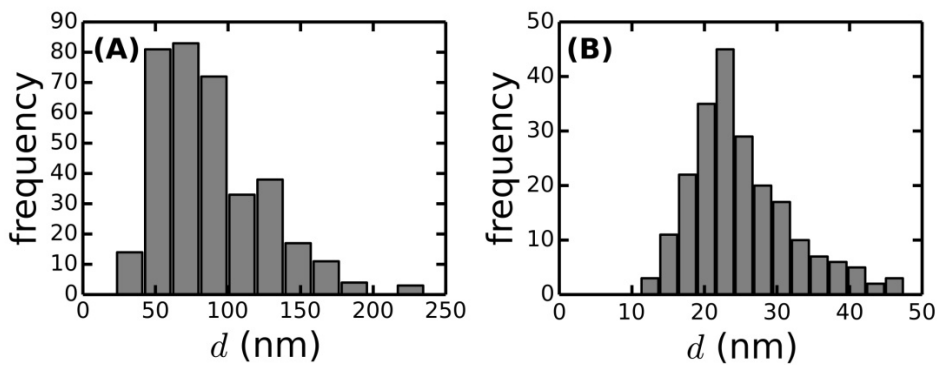

**Figure S1**. Histogram of fiber diameters determined from TEM images of fibers prepared under (A) coarse network conditions, which promote protofibril bundling, and (B) fine network conditions, which suppress protofibril bundling. In both cases, more than 200 fibers were taken into account, and data from networks polymerized at concentrations between 0.5 and 2 mg/ml were combined.

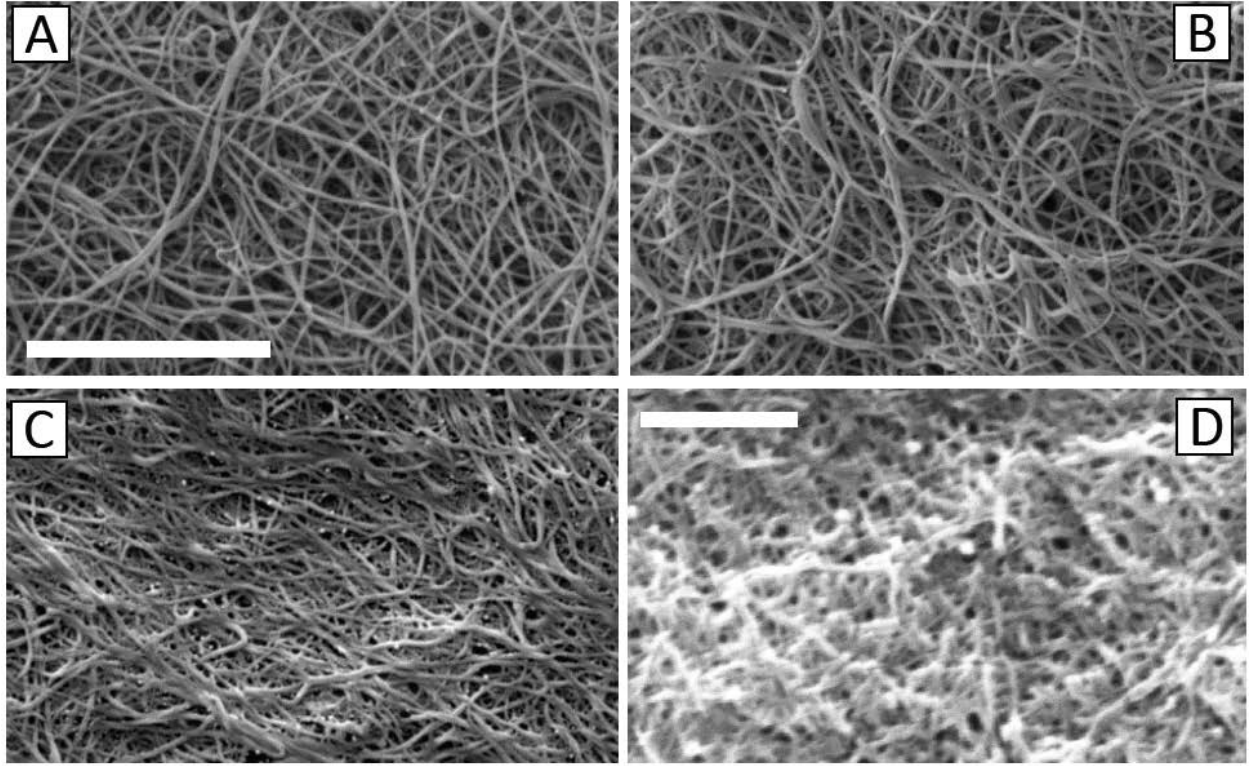

**Figure S2**. SEM images of fibrin networks. (A-C) Coarse fibrin networks polymerized at concentrations of 1, 3 and 7 mg/ml, respectively. Scale bar represents 5 µm for (A-C). (D) Fine fibrin network polymerized at a concentration of 1 mg/ml. Scale bar denotes 400 nm.

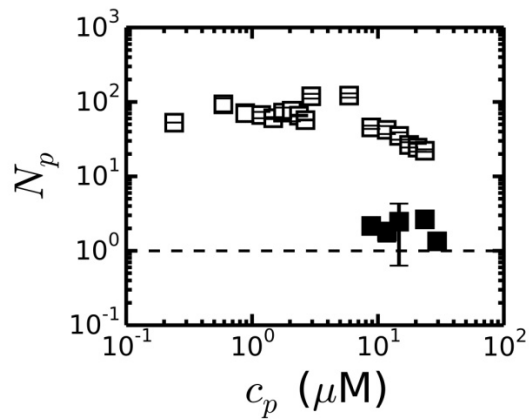

**Figure S3**. Number of protofibrils per fiber for fibrin fibers polymerized under coarse (open squares) and fine (filled squares) network conditions, based on turbidity measurements. The protofibril limit (i.e. $N_p$ = 1) is indicated.

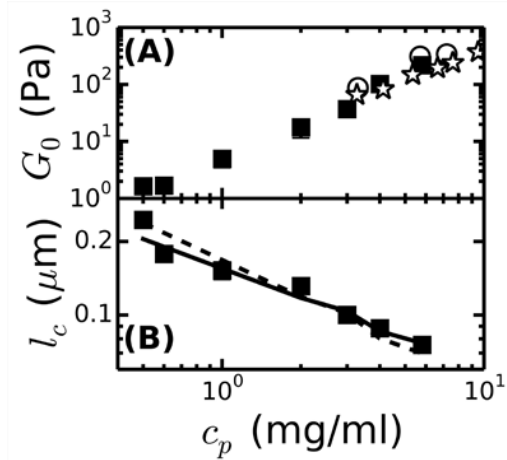

**Figure S4**. (A) The linear elastic modulus for fibrin clots polymerized under fine clot conditions, which show minimal bundling (black squares), compared with previous measurements (open circles: Ref. (1) and open stars: Ref. (2)). (B) Cross-link distance inferred by fitting the rheology data for fine clots to the affine model for wormlike chains to theoretical predictions according to $l_c \sim l_e = l_p^{1/5} \rho^{-2/5}$ (solid line) or $l_c \propto \rho^{-1/2}$ (dashed line), using a prefactor of 0.75 in both cases.

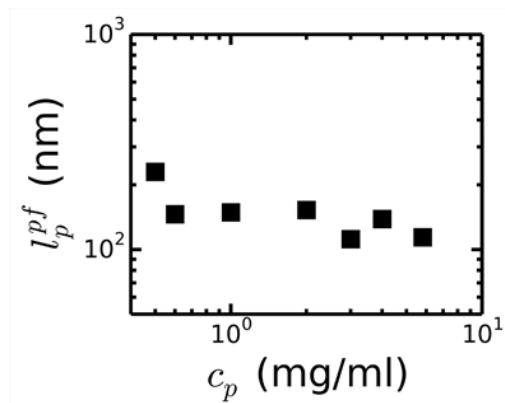

**Figure S5**. The persistence length of fibrin protofibrils, $l_p^{pf}$, obtained by fitting the full theoretical prediction for the stress-stiffening response of extensible wormlike chains to the fine fibrin rheology data. The persistence length does not vary significantly with concentration.

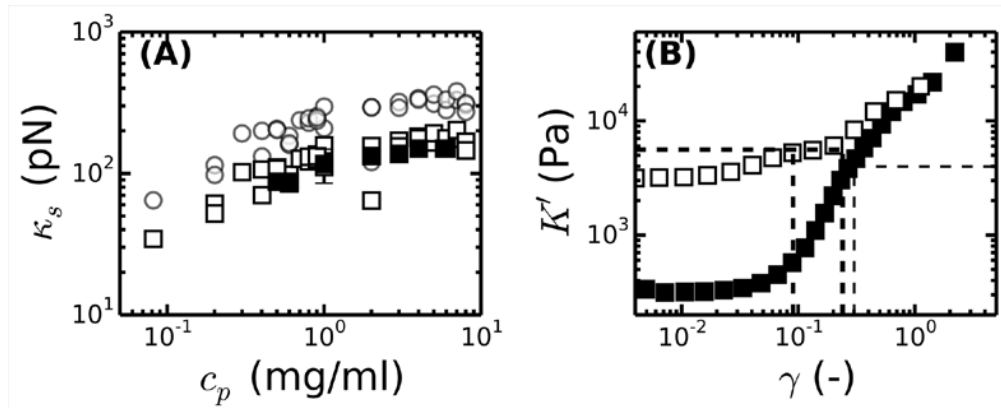

**Figure S6.** (A) The protofibril stretch modulus extracted from coarse and fine fibrin network rheology data. Solid squares represent data for fine fibrin, open squares represent data for coarse fibrin in the aligned limit ($K_s = 1/15\, \rho\, \kappa_s$), and open gray circles represent coarse clots in the isotropic limit ($K_s = 1/8\, \rho\, \kappa_s$). (B) The differential elastic modulus for 8 mg/ml fine (closed black squares) and coarse fibrin networks (open black squares) plotted against shear strain. For coarse fibrin, the region where $\kappa_s$ is determined is indicated in dashed lines. For fine fibrin, the inflection point is indicated by dashed lines, indicating the beginning of the enthalpic stretching regime. The stretch modulus is determined by fitting the non-linear mechanical properties by the full theoretical prediction (see Fig. 2A).

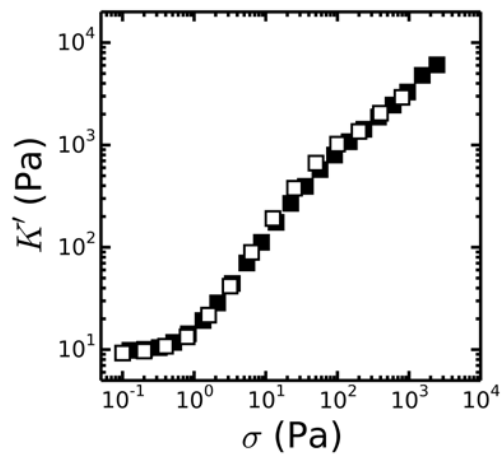

**Figure S7.** Nonlinear rheology of fine fibrin networks (2 mg/ml), in the presence (open squares) and absence (closed squares) of 200 µM D004, with 1% DMSO.

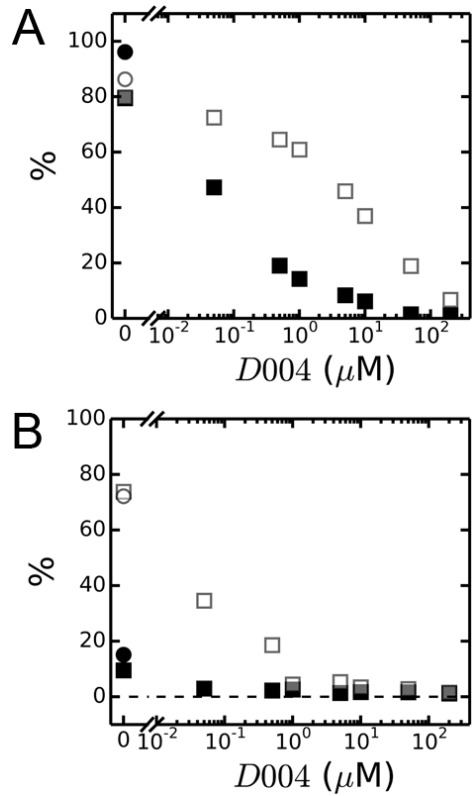

**Figure S8**. Percentage of crosslinked α-chains (closed black squares) and crosslinked γ-chains (open grey squares) in 2 mg/ml (A) coarse and (B) fine fibrin networks in the presence of varying amounts of FXIII inhibitor D004, determined by densitometric analysis of SDS-PAGE gel. Circles correspond to the 2 mg/ml fibrin control with no DMSO present.

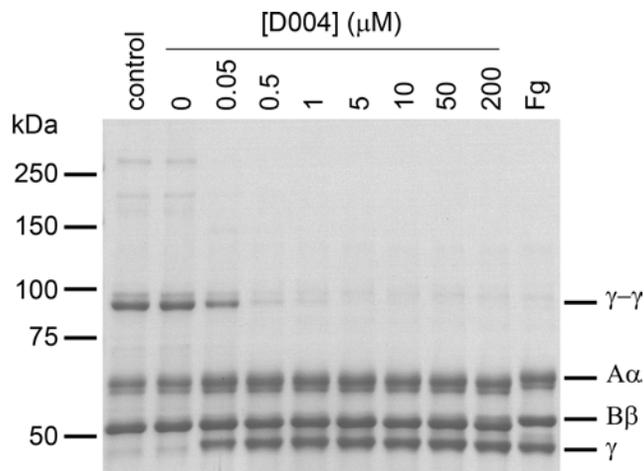

**Figure S9.** Reducing SDS-PAGE gel for 2 mg/ml fine fibrin networks formed in the presence of different concentrations of the FXIII inhibitor D004, as indicated. The control consists of fine fibrin without the presence of DMSO. Fg is fibrinogen in fine fibrin buffer without thrombin and calcium.

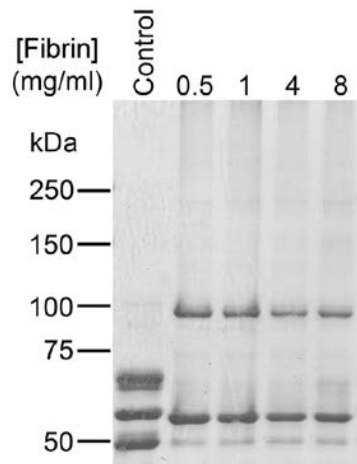

**Figure S10.** SDS-PAGE analysis of crosslinking of coarse fibrin networks with increasing protein concentration in mg/ml, as indicated. Control consists of fibrinogen without thrombin and calcium.